\documentclass[10pt]{iopart}

\usepackage{graphicx}
\usepackage{multicol}
\usepackage{blindtext}
\usepackage{float}
\usepackage{wrapfig}
\usepackage{lipsum}

\usepackage{eurosym}
\usepackage{amssymb}
\usepackage{amsmath}
\usepackage{amsfonts}
\usepackage{graphicx}
\usepackage{color}
\usepackage{bm}
\usepackage[numbers,square,sort&compress]{natbib}

\usepackage{cleveref}

\begin{document}

\title{Elimination of spurious oscillations on photoemission spectra}
\author{Mart\'{\i}n Barlari$^{1},$ Diego G. Arb\'{o}$^{1,2,3},$ Mar\'{\i}a
Silvia Gravielle$^{1},$ and Dar\'{\i}o M. Mitnik$^{1}$}
\address{$^{1}$ Instituto de Astronomía y Física del Espacio - 
IAFE (CONICET-UBA), 1428, Buenos Aires, Argentina}
\address{$^{2}$ Universidad de Buenos Aires, Facultad de Ciencias Exactas y
Naturales, Departamento de F\'{\i}sica, 1428, Buenos Aires, Argentina}
\address{$^{3}$ Universidad de Buenos Aires, Ciclo B\'{a}sico Com\'{u}n, Buenos
Aires, Argentina.}

\begin{abstract}
We present a method for accurately computing transition probabilities in one-dimensional photoionization problems. Our approach involves solving the time-dependent Schrödinger equation and projecting its solution onto scattering states that satisfy the correct incoming or outgoing boundary conditions. Conventionally, the photoelectron emission spectrum is obtained by projecting the time-evolved wavefunction onto the stationary continuum eigenstates of the unperturbed, time-independent Hamiltonian.
However, when the spatial potential is symmetric, both the initial bound state and the final continuum states exhibit well-defined parity. The propagated wavefunction retains structural features of the initial bound state, including its parity. As a result, changes in the parity of the continuum states can introduce substantial variations in the projections, leading to spurious oscillations in the computed electron emission spectrum.
Our method circumvents this issue by employing scattering states without defined parity. Furthermore, it enables the calculation of directional emission, making it possible to study emission asymmetries.
To illustrate the capabilities of our scattering projection method, we analyze the partial differential photoionization probabilities of Al(111) metallic surfaces under short laser pulses at grazing incidence.
\end{abstract}

\date{\today}

\maketitle

\section{{\protect\LARGE \protect\bigskip }Introduction}

Many ionization experiments aimed at obtaining energy spectra and cross sections are performed under quasi-stationary conditions. Consequently, quantum mechanics and scattering theory have traditionally focused on solutions of the time-independent Schr\"{o}dinger equation, primarily because stationary scattering states form a basis that facilitates the analysis of time-dependent collisions. However, recent breakthroughs in laser technology—particularly the advent of attosecond pulses, a key achievement recognized by the 2023 Nobel Prize in Physics \cite{Nobel2023}—have sparked growing interest in ultrashort laser pulses as ionizing sources. As a result, theoretical approaches must adapt to these developments by placing greater emphasis on the temporal description of physical processes.

The calculation of photoionization spectra has been an active area of research for several decades, encompassing a wide variety of targets, including atoms \cite{Agostini79,DiMauro95,Schultze10,Krausz09}, molecules \cite{Dutoi11,Huppert16,Ning14}, and solids \cite{Cavalieri07,Neppl12,Ossiander18}. In many cases—such as electron emission from solid surfaces—the underlying physics can be effectively described using one-dimensional models \cite{Rios12,Rios17}. For instance, photoemission from the valence band has been studied for metallic surfaces like Al(111), Al(100), Be(0001), and Mg(0001) \cite{Rios12,Rios17}.

Following previous studies \cite{Faraggi07,Rios12,Rios17,Fetic20}, photoelectron spectra are commonly obtained by numerically solving the time-dependent Schrödinger equation (TDSE) and projecting the final wavefunction onto the stationary continuum eigenstates of the unperturbed time-independent Hamiltonian. However, this approach often yields large unphysical oscillations in the energy spectrum \cite{Faraggi07,Rios12}. To mitigate these artifacts, convolution techniques such as the widely used window-operator method (WOM) \cite{Schafer90,Schafer91} are typically employed. Nevertheless, we have found that these methods can inadvertently suppress physically meaningful structures in the spectrum.

In this work, we restrict our study to one-dimensional ionization processes and perform a detailed analysis of electron emission from symmetric potentials. To overcome the aforementioned spurious oscillations, we introduce a computational method that projects the wavefunction—evaluated at the end of the laser pulse—onto scattering states with the appropriate incoming or outgoing boundary conditions. Our Scattering Projection Method (SPM) not only eliminates the unphysical oscillations more effectively than the WOM but also preserves fine structures of physical relevance, such as Ramsauer-Townsend-type oscillations, which are otherwise suppressed by the WOM.

We investigate the ionization of a one-electron system subjected to an external electric field, modeled as either a traveling wave (short laser pulse) or a standing wave (half-cycle pulse). We begin with a simple square well potential supporting a single bound state, then proceed to the jellium model \cite{Faraggi07} as a first approximation of a metallic surface, and finally apply a more sophisticated band-structure-based (BSB) potential for Al(111), which incorporates surface roughness due to atomic layering as well as surface plasmon effects \cite{Rios17,Chulkov99}.

This article is organized as follows. In Sec. \ref{theory}, we describe the numerical methods used to solve the TDSE for electron ionization in a one-dimensional potential, and we detail two approaches to extract the energy spectrum: (i) projection onto standard stationary continuum eigenstates of the time-independent Schrödinger equation (TISE), and (ii) the proposed SPM using scattering states. A step-by-step numerical procedure is provided in the Appendix. In Sec. \ref{resultsI}, we present and compare results for simple potentials obtained via both projection techniques and the WOM. In Sec. \ref{resultsII}, we apply the methods to the photoionization of metallic surfaces. Finally, Sec. \ref{conclusions} summarizes our findings and outlines future directions for the systematic study of metal photoionization under ultrashort laser pulses. Unless otherwise stated, atomic units are used throughout.

\section{Theory}

\label{theory}

\subsection{Resolution of the time-dependent Schr\"{o}dinger equation}

We consider the ionization of a one-dimensional system consisting of a single electron bound in a short-range potential $V(z)$ and subjected to an external time-dependent electric field. Within the dipole approximation, the time-dependent Schr\"{o}dinger equation (TDSE) takes the form
\begin{equation}
i \frac{\partial}{\partial t} |\psi(t)\rangle = \left[ H_0 + H_{\text{int}}(t) \right] |\psi(t)\rangle,
\label{TDSE}
\end{equation}
where the unperturbed Hamiltonian is $H_0 = p^2/2 + V(z)$, with the first term representing the kinetic energy of the electron, and the second term describing its interaction with an attractive potential that models, for example, an atom, molecule, or solid-state system.

The term $H_{\text{int}}(t)$ accounts for the interaction with the external field. In the length gauge, this interaction is given by $H_{\text{int}}(t) = z F(t)$, where $F(t)$ is the time-dependent electric field. Under the influence of this field, an electron initially in a bound state $|\phi_i\rangle$ may either be excited to another bound state $|\phi_n\rangle$ with energy $E_n$, or ionized into the continuum, ending in a state $|\phi_k\rangle$ characterized by asymptotic momentum $k$ and energy $E = k^2/2$.

To numerically solve Eq.~(\ref{TDSE}), we use the staggered leap-frog method \cite{Moler03, Press07}. This is a robust time propagation technique commonly applied in the simulation of atomic and molecular few-body dynamics \cite{Pindzola07}. The method provides an efficient and stable approach for evolving the wavefunction under the influence of strong and time-dependent external fields.

\subsection{Projection onto stationary waves}

Once the TDSE [Eq.~(\ref{TDSE})] has been solved, the time-propagated wave function $\psi (z,t)= \langle z|\psi (t) \rangle$ is available.
After the conclusion of the electric pulse of duration $\tau $, the electron
is subject only to the influence of the time-independent potential $V(z)$
Therefore, the kinetic energy is conserved, and the asymptotic momentum $k$ is a good quantum number.  
We consider in this paper only symmetric short-range potentials, i.e., 
$V(z)=V(-z)$.
Hence, both bound $\phi _{j}(z)$ and continuum $\phi _{k}(z)$ eigenstates of 
the time-independent Hamiltonian $H_{0}$
have definite parity, which means that they will be either even $[\phi _{i}^{(e)}(z)=\phi _{i}^{(e)}(-z)]$ or odd $[\phi_{i}^{(o)}(z)=-\phi _{i}^{(o)}(-z)]$. 
The wave function $\psi(z,\tau )$ at the end of the interaction with the external force can be written as a linear combination of stationary eigenstates
\begin{eqnarray}
\vert \psi(\tau) \rangle &=& \sum_{j}^{N_{b}}a_{j} \vert \phi_{j} \rangle +
\int_{-\infty }^{\infty }dk \, a_{k} \vert \phi_{k} \rangle \, ,
\label{wf}
\end{eqnarray}
where $N_{b}$ is the number of bound states of $V(z)$, and the transition amplitudes for bound and continuum states are 
\begin{eqnarray}
a_{i} &=& \langle \phi _{i} \, \vert \,    \psi (\tau ) \rangle  \, .
\label{ai}
\end{eqnarray}
The differential probability of electron emission can be expressed as a function of the electron momentum $k$ or the kinetic energy $E$ as 
\begin{equation}
\frac{dP}{dE}=k\frac{dP}{dk}= \sqrt{2E} \, \vert a_{k}\vert^{2}
\label{dPdE} \, .
\end{equation}
If the continuum states are degenerate, the photoelectron emission probability is
\begin{equation}
\frac{dP}{dE}=k\frac{dP}{dk} =\sqrt{2E} \left( 
\vert a_{k}^{(e)} \vert ^{2} + \vert a_{k}^{(o)} \vert ^{2} \right) \, ,
\label{spectrum-ideal}
\end{equation}
where the amplitudes $a_{k}^{(p)}$ correspond to the projections over the degenerate eigenfunctions 
$H_0 \vert \phi_k^{(p)}\rangle = E_k \vert \phi_k^{(p)} \rangle$, 
for $p=e,o$.

If the TISE is solved numerically within a bounded (1D) spatial range  (referred to as ``the box"), the Hamiltonian spectrum becomes $L^2$ finite and discrete even in the region where $(E > 0$).
The differential photoelectron emission probability can be recovered as
\begin{equation}
\frac{dP}{dE} \simeq \frac{|a_{n}|^{2}}{\Delta E_{n}} \qquad \mathrm{if} \quad  E_{n}>0 \, ,
\label{spectrum-box}
\end{equation}
where we have approximated the differential $dE$ for a discrete energy bin of size $\Delta E_{n}$, which takes into account the density of states. 
In this context, the index of the eigenfunction $n$ determines the parity of the eigenfunction.
From Eq. (\ref{spectrum-box}), we can express the total ionization probability $P_{\mathrm{ion}}$ as follows:
\begin{equation}
P_{\mathrm{ion}}=\int_{0}^{\infty }dE\left( \frac{dP}{dE}\right) \simeq
\sum_{n=N_{b}+1}^{N}|a_{n}|^{2},
\end{equation}
where the continuous integral over the energy was approximated for a discrete sum involving the $N-N_{b}$ eigenstates with $E_{n}>0$.

\subsection{Scattering projection method}

We can alternatively project the final state \(\left\vert \psi(t) \right\rangle\) onto a different basis of continuous states, which have non-definite parity. In this work, we propose the Scattering Projection Method (SPM), in which the projection basis consists of scattering waves that meet the appropriate asymptotic conditions. Given that the one-dimensional potential well is a short-range potential, the scattering waves with incoming (+) boundary conditions—related to plane-wave incidence from the right (r) and from the left (l)—asymptotically behave as follows:
\begin{subequations}
\begin{eqnarray}
\Psi _{k}^{(r,+)}(z) &=&\left\{ 
\begin{array}{lll}
T_{r}^{(+)}(k)e^{-ikz} & \mathrm{if} & z\rightarrow -\infty \\ 
e^{-ikz}+R_{r}^{(+)}(k)e^{ikz} & \mathrm{if} & z\rightarrow +\infty%
\end{array}%
\right.  \label{+r} \\
\Psi _{k}^{(l,+)}(z) &=&\left\{ 
\begin{array}{lll}
e^{ikz}+R_{l}^{(+)}(k)e^{-ikz} & \mathrm{if} & z\rightarrow -\infty \\ 
T_{l}^{(+)}(k)e^{ikz} & \mathrm{if} & z\rightarrow +\infty%
\end{array}%
\right.  \label{+l}
\end{eqnarray}
\end{subequations}

Similarly, the scattering waves with outgoing (-) boundary conditions, associated with the plane-wave emissions to the right (r) and the left (l), behave asymptotically as:

\begin{subequations}
\begin{eqnarray}
\Psi _{k}^{(r,-)}(z) &=&\left\{ 
\begin{array}{lll}
T_{r}^{(-)}(k)e^{ikz} & \mathrm{if} & z\rightarrow -\infty \\ 
e^{ikz}+R_{r}^{(-)}(k)e^{-ikz} & \mathrm{if} & z\rightarrow +\infty%
\end{array}%
\right.  \label{-r} \\
\Psi _{k}^{(l,-)}(z) &=&\left\{ 
\begin{array}{lll}
e^{-ikz}+R_{l}^{(-)}(k)e^{ikz} & \mathrm{if} & z\rightarrow -\infty \\ 
T_{l}^{(-)}(k)e^{-ikz} & \mathrm{if} & z\rightarrow +\infty%
\end{array}%
\right. . 
 \label{-l}
\end{eqnarray}
\end{subequations}
As it is well-known, both the incoming basis $\{\Psi ^{(r,+)},\Psi ^{(l,+)}\}$ and the outgoing basis $\{\Psi ^{(r,-)},\Psi ^{(l,-)}\}$ can be used to describe the continuous states of the scattering problem. 


%

\begin{figure}[!htb]
\centering 
\includegraphics[width=0.6\columnwidth]{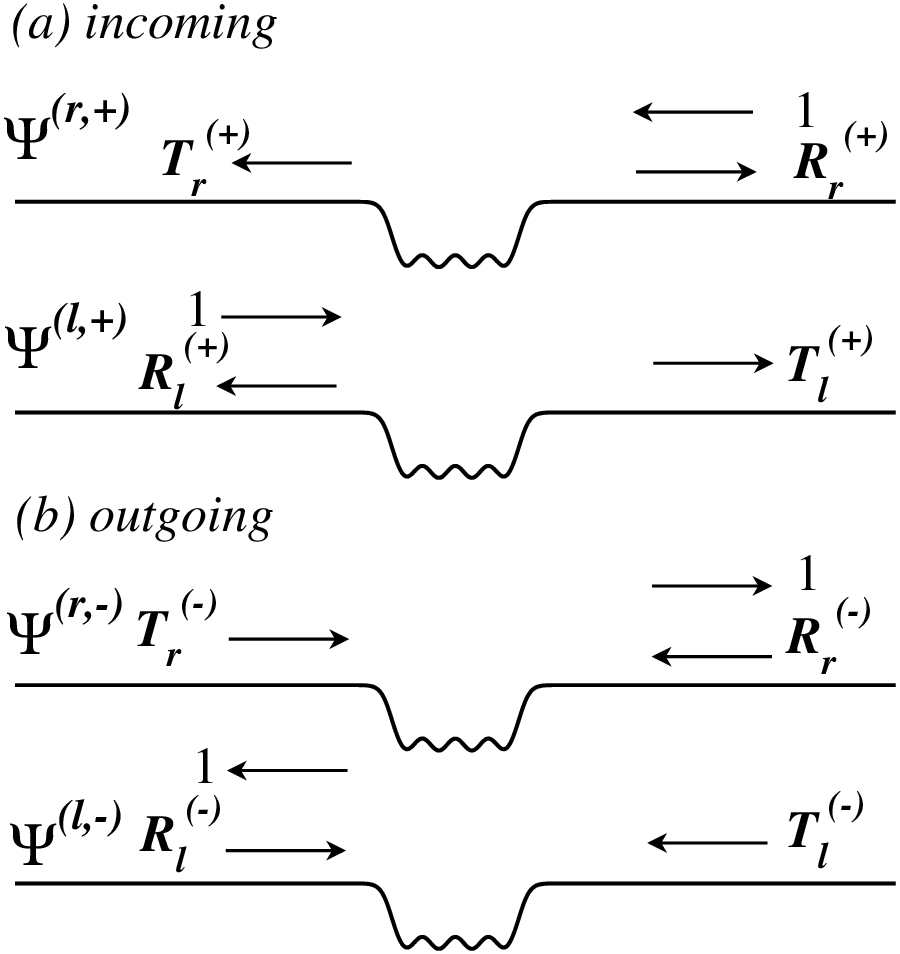} 
\caption{Scheme of scattering waves in a one-dimensional well.
(a) Incoming scattering waves from the right $\Psi _{k}^{(r,-)}$ and from the left $\Psi _{k}^{(l,-)}$ and (b) outgoing scattering waves to the right $\Psi_{k}^{(r,+)}$ and to the left $\Psi _{k}^{(l,+)}$, with their corresponding
reflection $R$ and transmission $T$ coefficients.}
\label{fig:esquema}
\end{figure}

In the specific case of a square-well potential, the scattering wave functions can be derived analytically \cite{Griffiths18}. For other potentials where the analytical scattering wave functions are either challenging to obtain or do not exist, we introduce a plane-wave basis using a procedure based on Ref. \cite{Pang06}. In \ref{appendixA}, we demonstrate how to numerically calculate the transmission and reflection coefficients for a given short-range potential. Once the scattering states are established, the photoemission probabilities can be retrieved within the SPM framework by projecting the time-propagated wave function onto the incoming (+) or outgoing (-) basis, similar to the process outlined in Eq. (\ref{spectrum-ideal}). That is  \cite{Garriz10},
\begin{subequations}
\begin{eqnarray}
\left( \frac{dP}{dE}\right) _{\mathrm{SPM}}^{(+)}=\sqrt{2E} &&\left( |\langle \Psi
_{k}^{(l,+)}|\psi (\tau )\rangle |^{2}\right.   \notag \\
&&\left. +|\langle \Psi _{k}^{(r,+)}|\psi (\tau )\rangle |^{2}\right) ,
\label{proba+} \\
\left( \frac{dP}{dE}\right) _{\mathrm{SPM}}^{(-)}=\sqrt{2E} &&\left( |\langle \Psi
_{k}^{(l,-)}|\psi (\tau )\rangle |^{2}\right.   \notag \\
&&\left. +|\langle \Psi _{k}^{(r,-)}|\psi (\tau )\rangle |^{2}\right) ,
\label{proba-}
\end{eqnarray}
\end{subequations}
respectively, and both expressions are equivalent.

\subsection{Window-operator method}

We provide a brief overview of the widely used window-operator method (WOM), developed by Schafer \cite{Schafer90,Schafer91}, for extracting energy-resolved probabilities from a wave function calculated on a numerical grid. This method approximates the differential emission probability by evaluating the expectation value of the window (or tapering) operator on the time-evolved wave function \(\left\vert \psi (\tau )\right\rangle\):
\begin{equation}
\left( \frac{dP}{dE}\right) _{\mathrm{W}}
\approx \langle \psi(\tau) | \hat{W} | \psi(\tau) \rangle =   
\langle \psi(\tau) | \frac{\gamma ^{2^{n}}}{\left( H_{0}-E\right) ^{2^{n}}+\gamma
^{2^{n}}} | \psi (\tau ) \rangle \, , 
\label{WOM1}
\end{equation}
which represents the probability of finding a particle within an energy bin of width \(2\gamma\), centered at the energy \(E\). 
Expanding the final numerical state into stationary wavefunctions,   Eq.~(\ref{WOM1}) becomes:
\begin{equation}
\left( \frac{dP}{dE}\right) _{\mathrm{W}}=\int\limits_{-\infty }^{\infty
}dk\frac{k\ \gamma ^{2^{n}}}{\left( \frac{k^{2}}{2}-E\right) ^{2^{n}}+\gamma
^{2^{n}}}\left\vert \left\langle \phi _{k}\right\vert \left. \psi (\tau
)\right\rangle \right\vert ^{2} \, ,    
\label{WOM2}
\end{equation}
which can be interpreted as the convolution of the amplitude \(\vert \langle \phi_{k} \vert \psi(\tau)\rangle \vert^{2}\) with the window operator function. 
In Eq. (\ref{WOM2}), we deal with eigenstates $\left\vert \phi _{k}\right\rangle $ in the continuum of energy; however, both bound and continuum states can be treated using the WOM. The two parameters of the window function, \(\gamma\) and \(n\), must be adjusted to obtain resolved and accurate energy spectra. For example, \(n = 1\) corresponds to a Lorentzian window, and as \(n\) increases, the window approaches a rectangular shape, thereby reducing the overlap between adjacent energy bins.

\section{Results for simple models}
\label{resultsI}

We evaluate different projection methods by calculating the electron emission spectra resulting from photoionization of a target subjected to an external linearly polarized laser, whose time-dependent electric field is given by 

\begin{equation}
F(t)=F_{0}\sin ^{2}\left( \frac{\pi t}{\tau }\right) \cos \left( \omega
t\right) \qquad  t=(0,\tau) \, ,  
\label{field}
\end{equation}%
where $F_{0}$ indicates the maximal strength of the electric pulse of main frequency $\omega $ and total
duration $\tau$. 

In order to understand the origin of the spurious oscillations commonly observed in the theoretical photoionization spectra, we start considering a laser interacting with an electron confined within a square potential well. This potential has a width of $a=3$ a.u. and depth $V_{0}=0.5$. It features a single bound state $\phi_1$ with an energy of $E_{1}=-0.314$ a.u.. 
The laser pulse has a frequency of $\omega =0.8$ a.u., peak field strength of $F_{0}=0.5,$ and a duration of six cycles, i.e., $\tau =6\times 2\pi /\omega =47.12$ a.u.. 
We solve the TDSE as given by Eq. (\ref{TDSE}) in the length gauge, where  $H_{\text{int}}(t)=zF(t)$. 
The spatial grid for our calculations consists of 5000 points evenly distributed over $z=(-L/2,L/2)$ with $L=200$ a.u.. The stationary eigenfunctions are obtained  by diagonalizing the $H_0$ Hamiltonian using the LAPACK package \cite{Lapack10}. 
For the TDSE solution, we utilize the staggered leap-frog method \cite{Moler03, Press07}, propagating the wavefunctions over 60000 time steps with  
$\Delta t=8.3\times 10^{-4}$ a.u. until a time slightly greater than the pulse duration. Throughout the propagation, we continuously monitor the proper normalization of the wave function to ensure unitarity.

\subsection{Differential emission probabilities}

\begin{figure}[!htb]
\centering 
\includegraphics[width=0.7\columnwidth]{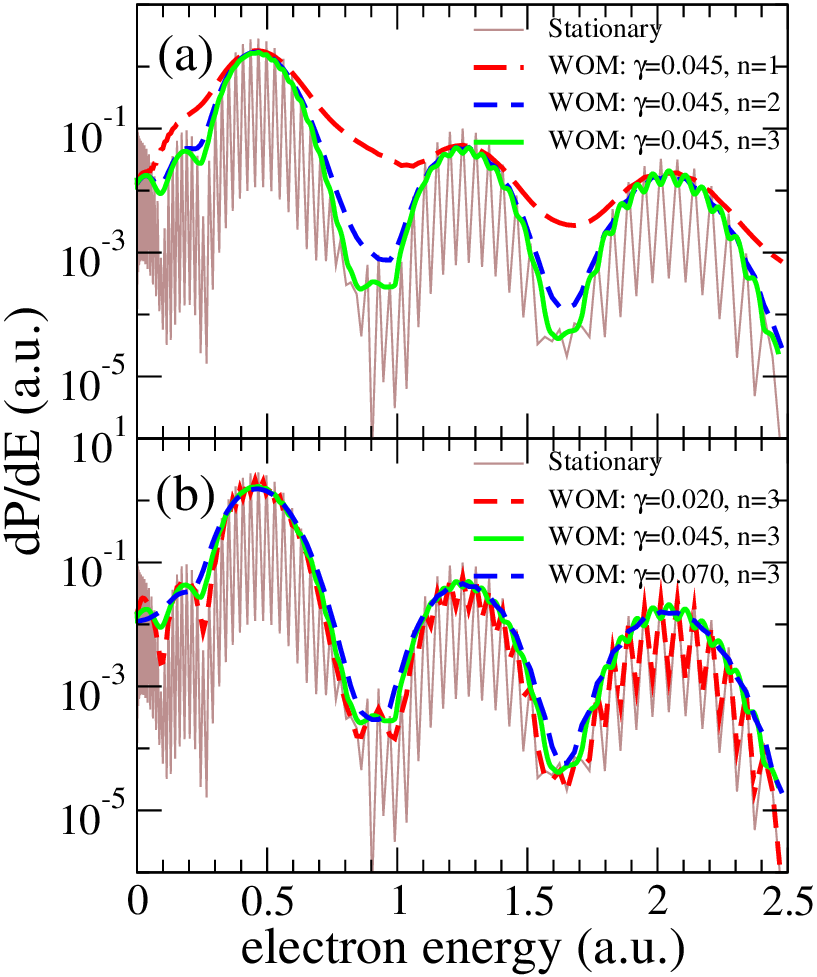} 
\caption{Ionization differential emission probabilities for an electron confined in a square well with $a=3$ a.u. and $V_{0}=0.5$ a.u. subject to a
laser pulse given by Eq.~(\ref{field}) with $\omega =0.8$ a.u., $F_{0}=0.5,$ and $\tau =47.12$ a.u.. 
(a) Projections onto stationary eigenstates (highly oscillating thin brown solid line) and WOM results with $\gamma =0.045$, for $n=1$, 2 (dash blue), and 3 (dash green).
(b) Projections onto stationary eigenfunctions (thin brown solid line), 
and WOM results with $n=3$ $\gamma =0.02$ (short dash blue), 0.045 (solid
green), and 0.07 (dash red) lines.}
\label{fig:schafer}
\end{figure}

The photoionization emission probability is usually calculated by
projecting the solution of the TDSE at the end of the laser pulse 
onto the continuum stationary eigenvectors. 
We computed the probabilities using Eq.~(\ref{spectrum-box}) and present the outcome in Fig.~\ref{fig:schafer}. 
The expected photoelectron emission spectra consists of broad peaks resulting from the multiphoton absorption process from the initial bound state. 
These peaks correspond to above-threshold ionization (ATI) energy levels ($n=1,2,\ldots$) and are separated by a photon energy of $\omega = 0.8$ a.u., in accordance with the energy conservation relation given by:
\begin{equation}
E_{n}=n\omega -I_{p}-U_{p} \, .
\end{equation}
In this context, the ionization potential is the binding energy of the initial state,  $I_{p}=|E_1|=0.314$ a.u., while the ponderomotive potential 
$U_{p}=\left( F_{0}/2\omega \right) ^{2}\simeq 0.1$ represents the energy of an oscillating electron influenced solely by the laser field \cite{Agostini79}. 

However, the resulting spectra exhibit a strong oscillating pattern superimposed on the expected curves. These rapid oscillations cover an amplitude of about two orders of magnitude, and the spacing between their peaks increases as the emitted electron energy increases. It becomes evident \cite{Garriz10} that the strong oscillatory structures within the peaks are a spurious artefact of the finite basis set used in the computations. The explanation for these unexpected oscillations is straightforward.  Since the time-evolved wavefunction $\psi(\tau)$ primarily retains the spatial shape of the initial bound state $\phi_1$, projecting onto continuum functions results in two types of transition amplitudes: one type, where both the initial $\phi_1$ (and consequently $\psi(\tau)$) and the projected $\phi_k$ waves have the same parity, retains nearly the entire probability.
The other type, in which the projected wavefunctions have opposite parities, exhibits significantly lower probabilities.
When the physical problem is confined to a numerical grid, it is possible to move arbitrarily the walls of the box, changing thus the stationary eigenfunctions. It is possible then to design two potentials in which the same eigenvalue appears in both cases, but the corresponding eigenvectors have different parities.
Therefore, the highly oscillating structures interfering with the ATI photoemission spectra are indeed spurious and should be disregarded.

\begin{figure}[!htb]
\centering 
\includegraphics[width=0.7\columnwidth]{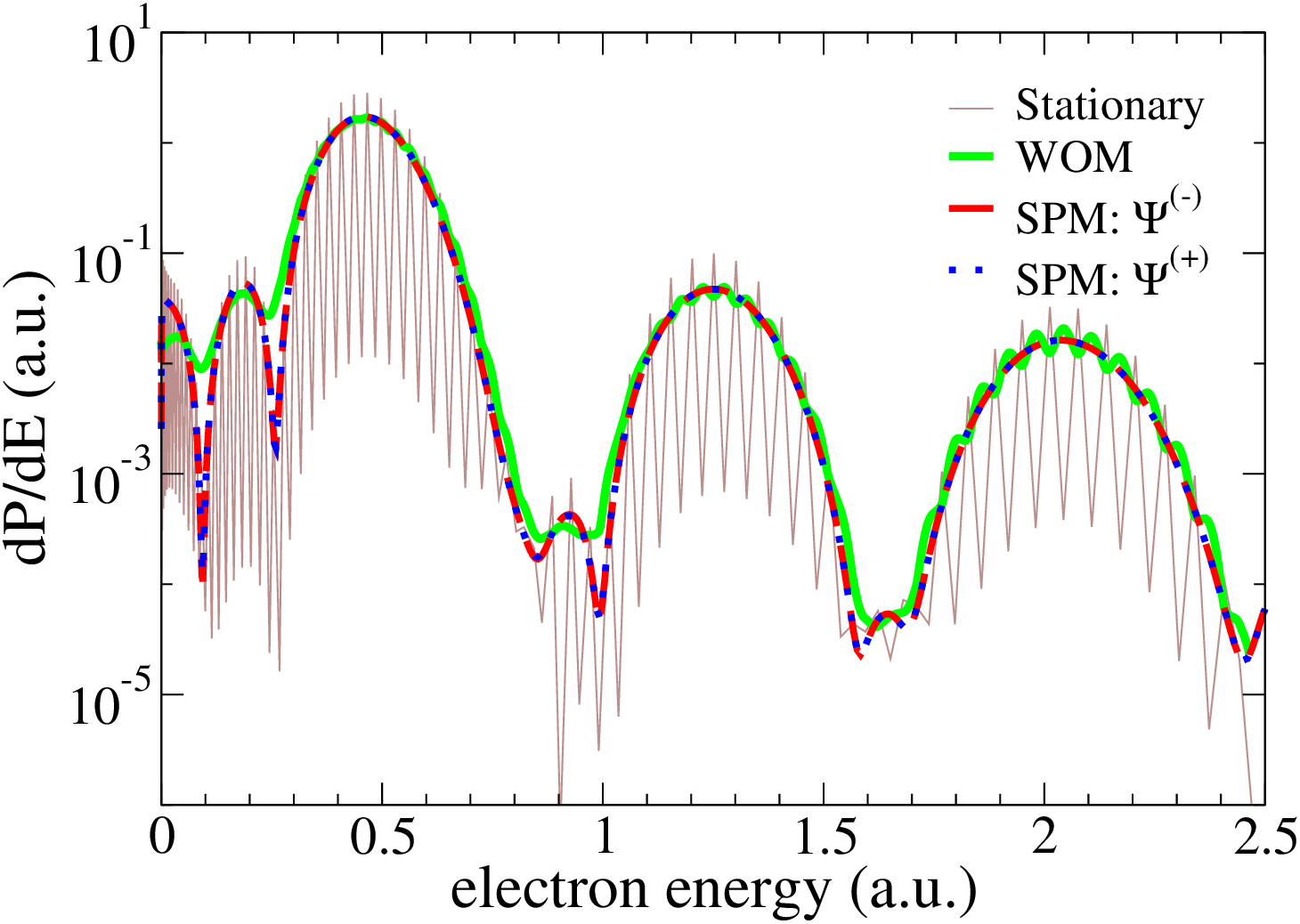} 
\caption{Photoelectron emission probabilities from $\phi_1$. 
Thin brown solid line: projection onto $L^2$ stationary eigenstates.
Dash blue line: $(dP/dE)_{\mathrm{SPM}}^{(+)}$, i.e., 
projection onto incoming $\Psi ^{(+)}$ scattering wave.
Solid red line: $(dP/dE)_{\mathrm{SPM}}^{(-)}$.
Green line: window filtered spectra with $\gamma =0.045$ and $n=$ $3$.}
\label{fig:jellium6}
\end{figure}

To reduce the high oscillations in the total energy spectrum, we initially employed the WOM method using Eq.~(\ref{WOM2}) with $\gamma =0.045$ and examined three different values of $n=1,2,$ and $3$, as illustrated in Fig. \ref{fig:schafer}a. On one hand, we find that the high-frequency fluctuations vanish within the WOM approach. Additionally, while the broad multiphoton peaks are somewhat smeared for $n=1$, they become well-defined with increasing values of $n$, with $n=3$ being sufficient for an acceptable resolution. In Fig. \ref{fig:schafer}b, we set $n=3$ and analyze the calculated spectrum for three distinct $\gamma$ values: 0.07, 0.045, and 0.02 a.u. 
The near-threshold structure, with peaks at approximately $E\simeq 0$ and 0.2 a.u., necessitates a narrow window with $\gamma =0.02$ a.u. However, at high energy, near the third multiphoton peak at $E\sim 2$ a.u., the spurious oscillations are not completely smoothed, even with this  $\gamma$ value. Thus, a compromise must be struck to find a suitable $\gamma$ value that addresses the near-threshold structures while mitigating the spurious oscillations at higher energies, which can sometimes prove challenging.
We then compare the outcome of the WOM with the electron emission spectrum obtained using our SPM. This was calculated by projecting the wave function at the end of the pulse, $\psi (z,\tau )$, onto the scattering waves $\Psi _{k}^{(l,\pm )}(z)$ and $\Psi _{k}^{(r,\pm )}(z)$, as described in Eqs.~(\ref{proba+}) and (\ref{proba-}). 
The photoelectron emission differential probability  
$\left(dP/dE\right) _{\mathrm{SPM}}^{(\pm )}$ calculated from those equations,  are shown in Fig. \ref{fig:jellium6}. As expected, both results are consistent with one another. Notably, utilizing the SPM yields a significantly better resolution of the physical structures across the entire energy spectrum compared to the WOM approach. For instance, the near-threshold structures and the secondary peaks between the multiphoton peaks at approximately $E \simeq 0.9$ a.u. and 1.65 a.u. are much more clearly resolved. In addition, the small oscillations still visible in the second and third multiphoton peaks, reminiscent of the spurious high-frequency peaks within the WOM, are entirely washed out within the SPM.

\subsection{Directional emission}

So far, we have analyzed the total emission spectrum without considering the emission direction. In one dimension, photoelectrons can be emitted either to the right (forward, along the polarization direction $\hat{\mathbf{z}}$) or to the left (backward). To compute directional emission, we project the final wavefunction onto the outgoing scattering states $\Psi_k^{(r,-)}(z)$ for emission to the right and $\Psi_k^{(l,-)}(z)$ for emission to the left:\begin{subequations}
\begin{eqnarray}
\left( \frac{dP}{dE}\right) _{r} &=&\sqrt{2E}|\langle \Psi _{k}^{(r,-)}|\psi
(\tau )\rangle |^{2}  \label{spectrum-r} \\
\left( \frac{dP}{dE}\right) _{l} &=&\sqrt{2E}|\langle \Psi _{k}^{(l,-)}|\psi
(\tau )\rangle |^{2}.  
\label{spectrum-l}
\end{eqnarray}
For symmetric potentials $V(z)$, such as those considered here, Eqs.~(\ref{spectrum-r}) and (\ref{spectrum-l}) can be related through time-reversal symmetry: $\Psi_k^{(r,+)}(z) = \Psi_k^{(l,-)*}(-z)$ and $\Psi_k^{(l,+)}(z) = \Psi_k^{(r,-)*}(-z)$, where -z is the position coordinate reflection upon the potential center. It follows from Eqs.~(\ref{proba+}) and (\ref{proba-}) that the total spectrum is simply the sum of the right and left contributions: $(dP/dE) = (dP/dE)_r + (dP/dE)_l$. 

Figure~\ref{fig:jelliumlr} shows the spectra corresponding to rightward, leftward, and total emission for the same square well and laser pulse considered previously. First, we numerically verify that projecting onto the basis $\Psi_k^{(r,+)}$ and $\Psi_k^{(l,-)}$ yields equivalent results for rightward emission, and analogously, that $\Psi_k^{(l,+)}$ and $\Psi_k^{(r,-)}$ are equivalent for leftward emission. More importantly, we observe that while the left and right emission spectra are similar, they are not identical. As demonstrated in Ref.~\cite{Arbo08a}, strictly symmetric emission arises under symmetric pulses $F(t) = F(\tau - t)$ within the strong-field approximation (SFA). In our case, the pulse is symmetric and the potential is short-ranged, so the SFA provides an accurate description. The small deviations from perfect symmetry observed in Fig.~\ref{fig:jelliumlr} stem from the depletion of the ground state and the residual effect of the short-range potential on the escaping electron.\footnote{The SFA is exact only for zero-range potentials and in the absence of initial state depletion.} In this example, the total emission to the right (left) accounts for 49.88\% (50.12\%) of the total.

\begin{figure}[!htb]
\centering 
\includegraphics[width=0.7\columnwidth]{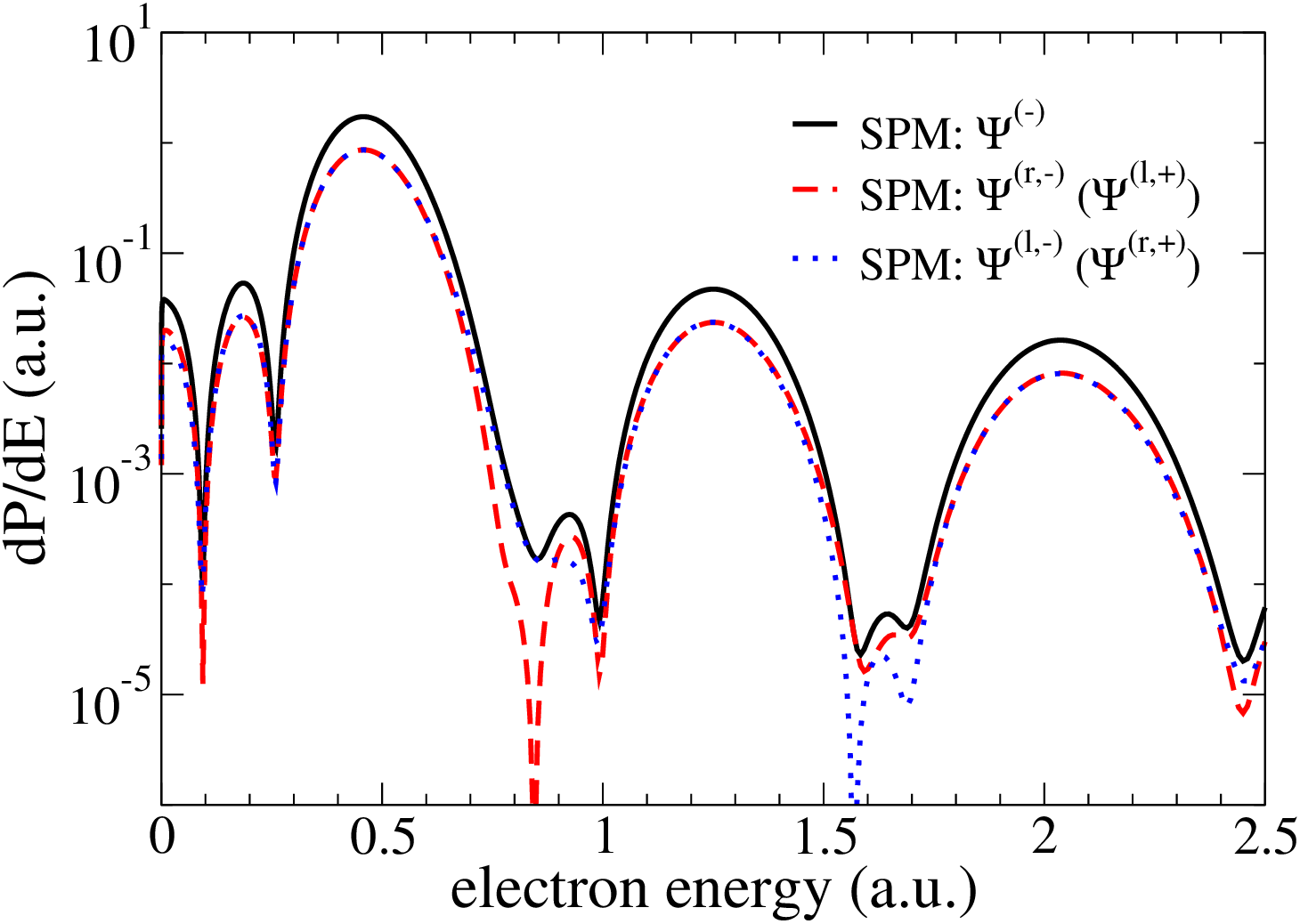} 
\caption{Projection of the photoelectron spectrum from the only
bound state of the same square well and subject to the same laser pulse as
in Figs. \ref{fig:schafer} and \ref{fig:jellium6} onto the scattering wave outgoing from the right $\Psi
^{(r,-)}$ or equivalently incoming to the left $\Psi ^{(l,+)}$ (see text) in
dashed red line, outgoing from the left $\Psi ^{(l,-)}$ or equivalently
incoming to the right $\Psi ^{(r,+)}$ in blue dotted line, and the sum of
the two, i.e., projecting onto $\Psi ^{(-)}$ or equivalently onto $\Psi
^{(+)}$ in black solid line.}
\label{fig:jelliumlr}
\end{figure}

Now, we consider photoionization due to a short laser pulse with just two
optical cycles, i.e., $\tau =2\times 2\pi /\omega =15.71$ a.u.. In Fig. \ref{fig:jellium2}a,
we plot the total energy spectrum calculated by projecting onto stationary
eigenfunctions of the box, resulting in a highly oscillating distribution. By
applying the WOM with $n=3$ and $\gamma =0.045$, we can smear the
spurious oscillations rather efficiently; however, we still observe a
reminiscence of these oscillations at high energy ($E\gtrsim 1.3$ a.u.).
In addition, in the low energy part of the spectrum near threshold ($E\lesssim
0.05$ a.u.), the WOM spectrum falls off abruptly. This is an artifact of the
method since it averages an energy region of width close to $2\gamma $,
whose one part lies in the continuum with appreciable transition
probabilities ($E>0$) and the other lies below the threshold ($E\,<0$) with
no eigenenergies close to it. Thus, the WOM average in Eq. (\ref{WOM2})
decreases as the energy approaches the threshold. Unlike the WOM, SPM
succeeds in losing completely any reminiscence of the spurious oscillations
in the energy spectrum and does not show any non-physical fall-off behavior
near threshold. For such a short pulse, the multiphoton peaks are absent in
the photoelectron spectrum because of the broad band frequency components of
the short laser field. Generally, as the laser pulse has a finite duration $%
\tau $, the uncertainty relation $\Delta E\Delta t\sim 1$ leads to
multiphoton peaks with a certain width $\Delta E\sim 1/\tau $. In this case,
as the laser pulse comprises only two cycles, $\tau $ is small and the
multiphoton peaks are very broad so that they overlap, which prevents their
visualization. In Fig. \ref{fig:jellium2}b, the electron emission to the right $\left(
dP/dE\right) _{r}$ and to the left $\left( dP/dE\right) _{l}$ are displayed.
For this such short laser pulse, we observe a rather asymmetric
distribution, except near threshold. In particular, the total emission to
the right (left) is $42.5\%$ ($57.5\%$) of the total ionization probability $%
P_{\mathrm{ion}}=0.64$ in this case. As expected, the right-left asymmetry
is enhanced when the pulse duration is shortened (compare Fig. \ref{fig:jellium2} to Fig. \ref{fig:jelliumlr}).

\begin{figure}[!htb]
\centering 
\includegraphics[width=0.7\columnwidth]{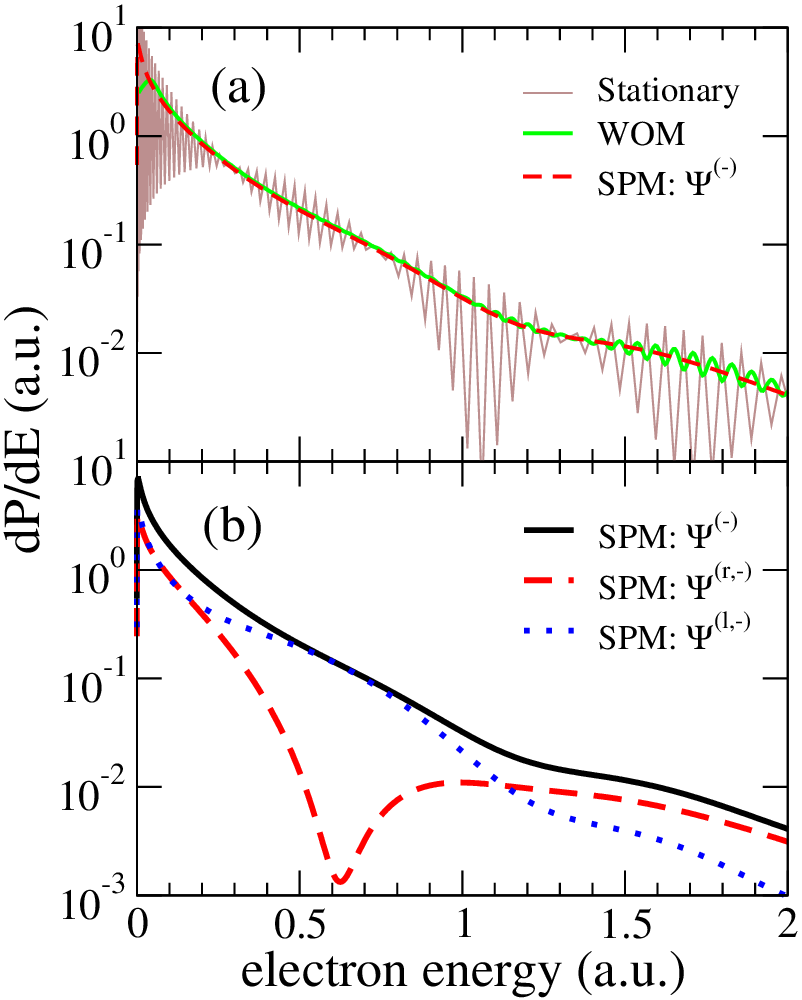} 
\caption{Photoelectron spectrum from the only bound state of the
same square well as in Figs. \ref{fig:schafer}-\ref{fig:jelliumlr}, subject to one laser pulse with the same
characteristics as in Figs. \ref{fig:schafer}-\ref{fig:jelliumlr} but with only two optical cycles of
duration, i.e., $\tau =15.71$ a.u.. (a) Projection of the spectrum onto
stationary states with definite parity in thin brown solid line (highly
oscillating), the spectrum projected onto the outgoing $\Psi ^{(-)}$ in red
dashed line and comparison to the spectrum calculated through the WOM with
window width $\gamma =0.045$ and $n=$ $3$ in green solid line. (b)
Directional photoelectron spectrum calculated projecting the final wave
function onto $\Psi ^{(r,-)}$ and $\Psi ^{(l,-)}$ together with their sum
(projection onto $\Psi ^{(-)}$). The ionization towards the right (left) is $%
42.5\%$ ($57\%$).}
\label{fig:jellium2}
\end{figure}

In order to produce very asymmetric electron emission, we consider
half-cycle pulses of the form 
\end{subequations}
\begin{equation}
F(t)=-F_{0}\sin ^{2}\left( \frac{\pi t}{\tau }\right) ,  \label{HCP}
\end{equation}%
for $0<t<\tau $ and zero elsewhere. These short electric fields of duration
in the order of the picoseconds do not represent traveling waves and have
been developed for the production of quasi-one-dimensional very-high-$n$
Rydberg atoms \cite{Lancaster03} and inducing focusing of Rydberg wave
packets in the phase space \cite{Arbo03}. However, in this work we will use
pulses much shorter than the excursion time of the initial state, i.e., $%
\tau \ll 2a/\sqrt{2I_{p}}$ (in the order of the attoseconds) to ensure that
we are in the sudden regime where the electric field of Eq. (\ref{HCP}) can
be thought of as a single kick of momentum transfer $\Delta
p=\int_{0}^{\infty }dtF(t)$ when they interact with the electron initially
bound to the square-well potential used. In Fig. \ref{fig:jelliumhalf} we show the directional
(right and left) emission spectrum from the same square well of Figs. \ref{fig:schafer}-\ref{fig:jellium2}
subject to the half-cycle pulse [Eq. (\ref{HCP})] of duration $\tau =1$ a.u.
and momentum transfers $\Delta p=-0.05$ a.u. and $\Delta p=-0.5$ a.u. in
Figs. \ref{fig:jelliumhalf}a and \ref{fig:jelliumhalf}b, respectively. We observe that there is more emission
towards the right than towards the left for both kick strengths, which is
anticipated from a classical viewpoint, since negative kicks push the
electron (of negative charge) towards the right. As expected, when the kick
strength is increased, the right-left emission asymmetry increases. For $%
\Delta p=-0.05,$ the emission to the right (left) is $57\%$ ($43\%$) of the
total ionization probability is $P_{\mathrm{ion}}=3\times 10^{-4},$ whereas
for $\Delta p=-0.5$, the emission to the right (left) is $92\%$ ($8\%$) of
the total ionization $P_{\mathrm{ion}}=0.27$.

\begin{figure}[!htb]
\centering 
\includegraphics[width=0.7\columnwidth]{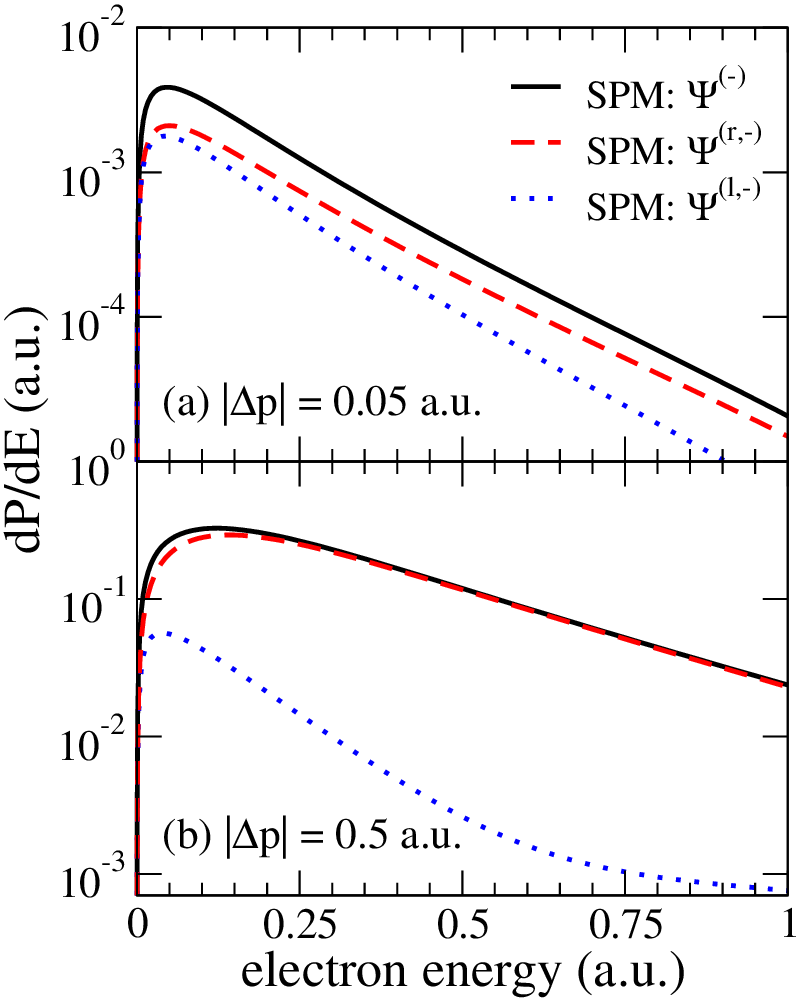} 
\caption{Photoelectron spectrum from the only bound state of the
same square well in Figs. \ref{fig:schafer}-\ref{fig:jellium2} subject to a half-cycle pulse of duration $%
\tau =1$ a.u. and momentum transfer $\Delta p=-0.05$ a.u. in (a) and $\Delta
p=-0.5$ a.u. in (b). The spectrum projected onto the respective incoming
scattering waves from right and the left $\Psi ^{(r,-)}$ and $\Psi ^{(l,-)}$
together with their sum (projection onto $\Psi ^{(-)}$).}
\label{fig:jelliumhalf}
\end{figure}

\begin{figure}[!htb]
\centering 
\includegraphics[width=0.7\columnwidth]{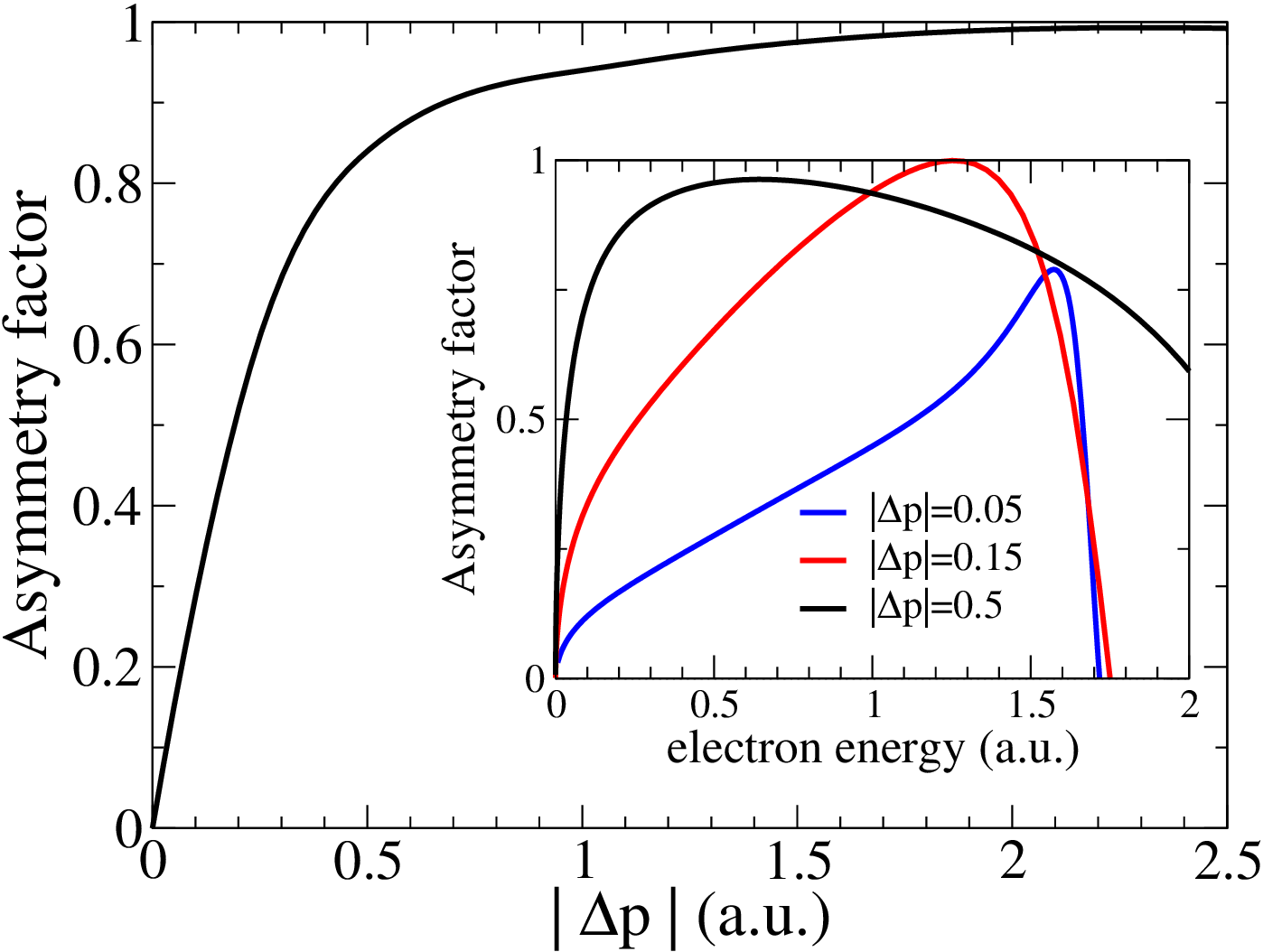} 
\caption{Asymmetry factor $A$ as a function of the magnitude of
the momentum transfer $\left\vert \Delta p\right\vert $ for ionization by a
half-cycle pulse of duration $\tau =1$ a.u.. Inset: differential asymmetry
factor $A(E)$ as a function of the energy for three different momentum
transfers: $\Delta p=-0.05,$ $-0.15,$ and $-0.5.$}
\label{fig:asym}
\end{figure}

For a closer inspection of the emission asymmetry for half-cycle pulses, we
define the total asymmetry factor as%
\begin{equation}
A=\frac{\left( P_{\mathrm{ion}}\right) _{r}-\left( P_{\mathrm{ion}}\right)
_{l}}{P_{\mathrm{ion}}},  \label{A}
\end{equation}%
where $\left( P_{\mathrm{ion}}\right) _{r}=\int_{0}^{\infty }dE\left(
dP/dE\right) _{r},$ $(P_{\mathrm{ion}})_{l}=\int_{0}^{\infty }dE\left(
dP/dE\right) _{l},$ and $P_{\mathrm{ion}}=\left( P_{\mathrm{ion}}\right)
_{r}+(P_{\mathrm{ion}})_{l},$ and the differential asymmetry factor as 
\begin{equation}
A(E)=\frac{\left( \frac{dP}{dE}\right) _{r}-\left( \frac{dP}{dE}\right) _{l}%
}{\left( \frac{dP}{dE}\right) }.  \label{AofE}
\end{equation}%
The asymmetry factors [Eq. (\ref{A})] for the right-left spectra of Fig. \ref{fig:jellium2}b,
Fig. \ref{fig:jelliumhalf}a, and Fig. \ref{fig:jelliumhalf}b are $A=-0.15,0.15,$ and $0.84$, respectively. In Fig.
\ref{fig:asym}, we show the asymmetry factor $A$ defined in Eq. (\ref{A}) as a function
of the magnitude of the momentum transfer $\left\vert \Delta p\right\vert $
for ionization by a half-cycle pulse of duration $\tau =1$ a.u.. We clearly
observe that the asymmetry factor increases monotonically as the momentum
transfer increases in magnitude, tending to unity, which means that all the
ionization will be towards the right and, on the other hand, for weak
momentum transfers the asymmetry approaches to zero, as expected \cite%
{Arbo03}. In the inset we show the differential asymmetry coefficient $A(E)$
defined in Eq. (\ref{AofE}) and observe that different momentum transfers
impacts differently on the emission asymmetry depending on the kinetic
energy of the escaping electron.

\section{Applications to metallic surfaces}

\label{resultsII}

In this section, we apply our study on the retrieval of the energy spectrum
and the elimination of spurious oscillations to photoemission from metallic
surfaces. We consider grazing incidence of an ultrashort laser pulse of
duration in the range from femto- to attoseconds on a metallic surface. We
characterize the pulse as a time-dependent electric field linearly polarized
along $\hat{\mathbf{z}},$ perpendicular to the solid surface within the
dipole approximation. Because of this interaction, the solid surface is
ionized, so that an electron located in the valence band of the solid is
promoted to the continuum. Using the single active electron (SAE)
approximation, the total Hamiltonian is described by

\begin{equation}
H(\mathbf{r},t)=-\frac{\nabla _{r}^{2}}{2}+V_{S}(z)+zF(t),  \label{eq:Hamilt}
\end{equation}

\noindent where $-\nabla _{r}^{2}/2$ is the electron kinetic energy, $%
V_{S}(z)$ is the potential that represents the electron-surface interaction
and $zF(t)$ accounts for the interaction of the active electron with the
electric field , within the length gauge. We consider a laser pulse of the
form of Eq. (\ref{field}). The Hamiltonian in Eq. (\ref{eq:Hamilt}) can be
separated in the Cartesian coordinates $%
H(x,y,z,t)=H_{x}(x)+H_{y}(y)+H_{z}(z,t)$ with $H_{x}(x)=-\partial
^{2}/\partial x^{2}$ and $H_{y}(y)=-\partial ^{2}/\partial y^{2}$
corresponding to the Hamiltonian of a free particle, and 
\begin{equation}
H_{z}(z,t)=-\partial ^{2}/\partial z^{2}+V_{S}(z)+zF(t),  \label{Hz}
\end{equation}%
corresponding to a one-dimensional model of the surface. The electron
dynamics is determined by the TDSE in Eq. (\ref{TDSE}). As the solutions of $%
H_{x}(x,t)$ and $H_{y}(y,t)$ are just plane waves, we circumscribe our
problem to solve the TDSE in the $\hat{\mathbf{z}}$ direction perpendicular
to the solid surface.

\subsection{Jellium surface model}

\begin{figure}[!htb]
\centering 
\includegraphics[width=0.7\columnwidth]{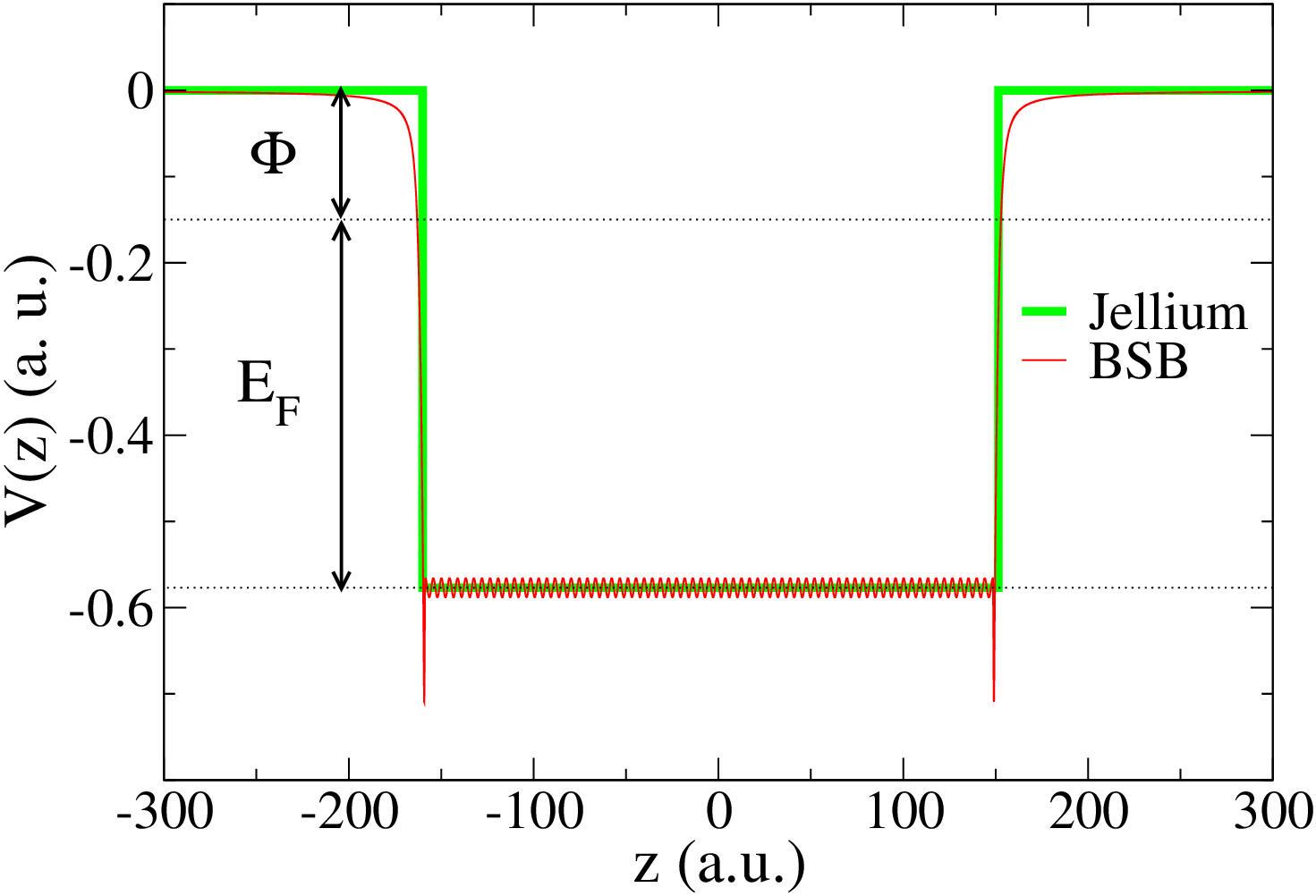} 
\caption{Jellium and BSB model potentials for the metallic surface
Al(111) with $E_{F}=0.414$ a.u., $\Phi =0.156$ a.u., and $a=311.54$
a.u.}
\label{fig:pot}
\end{figure}

First, we analyze photoemission for the simplest possible model for
the electron-surface interaction: the jellium model, where the electron-surface
interaction is  modeled as a slab of width $a$ and depth $%
V_{0}=E_{F}+\Phi ,$ corresponding to a Fermi energy $E_{F}$ and a work
function of the metal surface $\Phi ,$ thus

\begin{equation}
V_{S}(z)=-V_{0}\Theta (a/2-|z|).  \label{eq:jellium}
\end{equation}%
As the room temperature is much less than the Fermi temperature of metals,
the initial state corresponds to a set of electrons with energy levels below
the Fermi level (from the bottom of the potential well), i.e., $%
-V_{0}<E<-V_{0}-E_{F}=-\Phi ,$ due to the Pauli exclusion principle (see
Fig. \ref{fig:pot}). Therefore, we must calculate the photoelectron spectrum as \cite%
{Rios15} 
\begin{equation}
\frac{dP}{dE}=2\sqrt{2E}\rho (E)\sum_{i}|a_{ik}|^{2}\Theta (-E_{{i}}-\Phi ),
\label{spectrum-surface}
\end{equation}%
where the factor $2$ arises from spin independence and $\rho (E)$ is the
density of final continuum states with final electron momentum perpendicular
to the surface $k=\sqrt{2E}$. The Heaviside function $\Theta (-E_{i}-\Phi )$
restricts the initial states to those contained within the Fermi sphere
(reduced to an energy segment in our 1D treatment), where $E_{i}$ is the
initial bound electron energy.

\begin{figure}[!htb]
\centering 
\includegraphics[width=0.7\columnwidth]{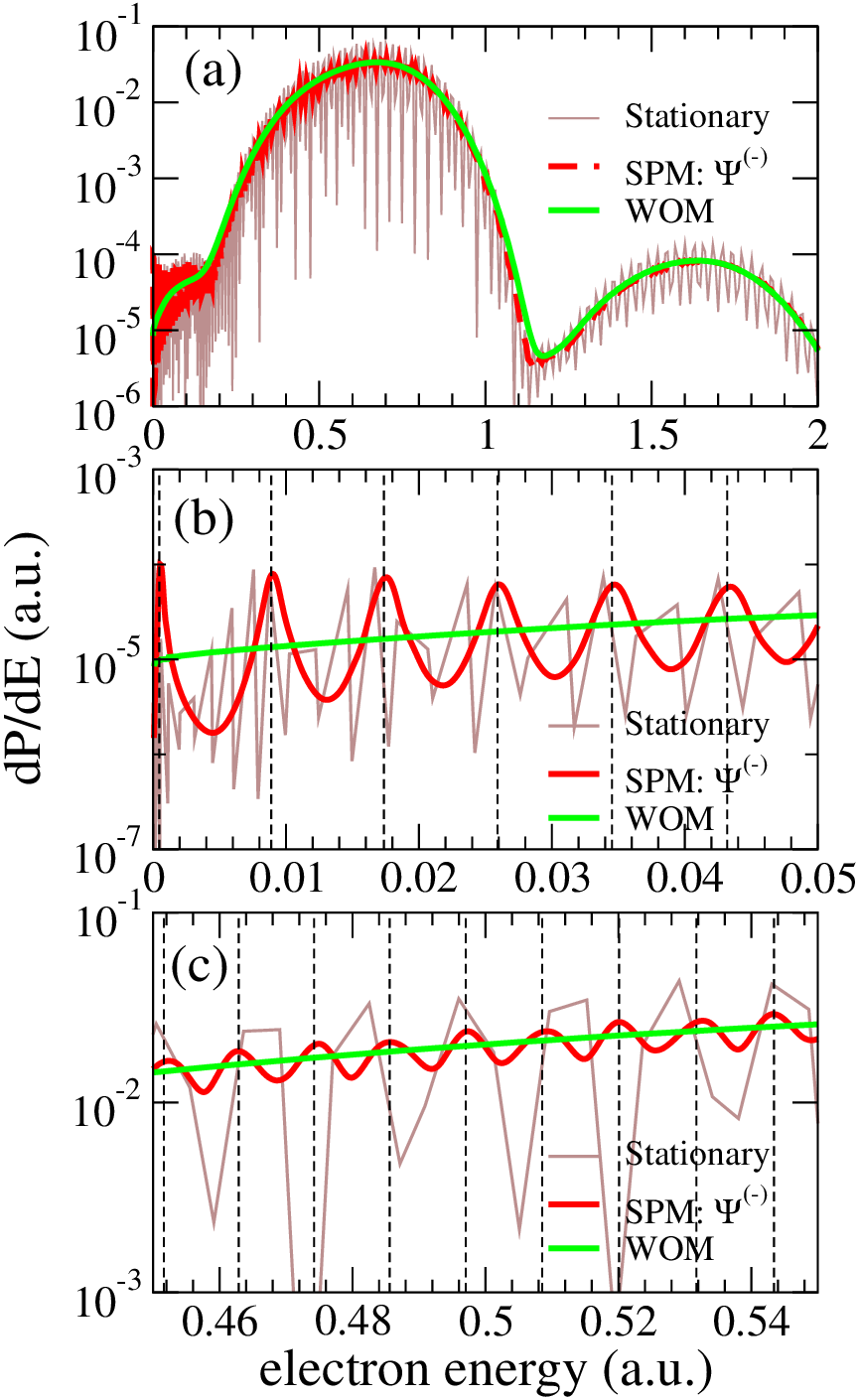} 
\caption{(a) Photoelectron spectrum perpendicular to the surface
Al (111) within the jellium model: $\Phi =0.156$ a.u., $E_{F}=0.414$ a.u., $%
a=400$ a.u., $V_{0}=0.57$ a.u.. Projection onto stationary states of
definite parity in brown thin solid line (highly oscillating) and projection
onto outgoing states $\Psi ^{(-)}$ in red dashed solid line and the WOM
spectrum in thick solid green line ($\gamma =0.045$ a.u. and $n=3$). Close
up for the ranges (b) $0<E<0.05$ near threshold and (c) $0.45<E<0.55$ near
the top of the first multiphoton peak. Vertical dash lines correspond to
Ramsauer-Townsend resonances, i.e., $E_{j}=j^{2}\pi ^{2}/(2a^{2})-V_{0}$
with $V_{0}=$ $\Phi +E_{F}.$ Laser pulse parameters are $F_{0}=0.1$ a.u.$,$ $%
\omega =1$ a.u., and $\tau =40$ a.u.}
\label{fig:aljellium}
\end{figure}

We solve the one-dimensional TDSE corresponding to the problem of
photoemission perpendicular to Al(111) surface for the jellium model Eq. (%
\ref{eq:jellium}) using $a=400$ a.u., and $V_{0}=E_{F}+\Phi =0.57$
corresponding to a Fermi energy $E_{F}=0.414$ a.u. and a work function of
the metal surface $\Phi =0.156$ a.u.. We consider the ionization of Al(111)
due to grazing incidence of a laser pulse of electric field amplitude $%
F_{0}=0.1$ a.u., frequency $\omega =1$ a.u. and duration equal to $\tau =40$
a.u.. By solving the TDSE via a finite difference method, we get the
corresponding time-evolved wave function at the end of the laser pulse $\psi
(z,\tau )$. Then, we project $\psi (z,\tau )$ onto the stationary continuous
eigenfunctions of the time-independent part of the Hamiltonian in Eq. (\ref%
{Hz}) and get the transition coefficients whose modulus squared represent
the emission probability of the electron into the vacuum region with energy $%
E=k^{2}/2$ [see Eq. (\ref{spectrum-box})]. In this case, we solve the TDSE
in Eq. (\ref{TDSE}) with the Hamiltonian of Eq. (\ref{Hz}) discretizing the
position with a numerical mesh of width $L=800$ a.u. and spacing $\Delta
z=0.1$ a.u.. We compute the time propagation up to the end of the laser
pulse $\tau $ using a time step $\Delta t=4.05\times 10^{-4}$ a.u.. The
aforementioned unperturbed Hamiltonian of Eq. (\ref{Hz}) with the jellium
potential in Eq. (\ref{eq:jellium}) is diagonalized \cite{Lapack10},
resulting in $136$ bound eigenstates, from which $116$ are initially
occupied below the Fermi level, and $507$ continuum eigenstates. Because the
jellium potential is even with respect to the center of the well,
eigenstates of the time-independent part of $H_{z}$ in Eq. (\ref{Hz}) are
even or odd \cite{Griffiths18}. In addition, the corresponding time-evolved
wave function for the typical laser intensities considered is only subtly
non-symmetrical. We then projected the time-evolved wave-function at the end
of the pulse, $\psi (z,\tau ),$ onto the stationary eigenstates of the
numerical box with eigenenergies $E>0$ (representing the continuum), which
results in the energy spectrum with huge oscillations spanning about two
orders of magnitude shown with a thin red line of Fig. \ref{fig:aljellium} as reported in
previous works \cite{Faraggi07,Rios12,Garriz10}. The corresponding projections
on the stationary eigenstates result in high or low values depending on
whether the parity of the time-evolved wave function matches or not the
parity of the continuum eigenstate, respectively. Consequently, we obtain
the very oscillating spectrum in Fig. \ref{fig:aljellium}a. We observe the multiphoton peaks
at $E\simeq 0.7$ a.u. and $1.7$ a.u. separated by the photon energy $\omega
=1$ a.u.. We use the WOM with width parameter $\gamma =0.045$ a.u. and $n=3$
to smooth the spectrum, which is observed in green solid line in the whole
energy range. SPM results also in a smooth spectrum in the high-energy
region but many oscillations arises for $E\lesssim 0.8$ a.u. The amplitude
of these high frequency oscillations increases substantially as the energies
approach the threshold. We show a close-up of Fig. \ref{fig:aljellium}a near threshold in Fig.
\ref{fig:aljellium}b and near the first multiphoton peak in Fig. \ref{fig:aljellium}c. We observe that the high
oscillations exposed by the SPM have a spacing which increases with the
electron kinetic energy and perfectly matches the eigenenergies of the
infinite well of width $a$ and depth $V_{0},$ indicated with vertical dashed
lines. This is a proof that these high-frequency oscillations are of the
different nature from the spurious high-frequency oscillations when
projecting on stationary eigenstates of the box with definite parity. The
oscillations captured by the SPM corresponds to Ramsauer-Townsend
oscillations coming from the interference of the travelling waves bouncing
against the sharp edges of the well with energies $E_{j}=j^{2}\pi
^{2}(2a^{2})-V_{0}$. We clearly see that the SPM accounts accurately for any
high-frequency physically meaningful oscillation in the energy spectrum
whereas the WOM overlooks such resonances. . However, if the potential well
described a finite system such as nanostructures or quantum dots,
Ramsauer-Towsend oscillations or any kind of high frequency oscillation do
have a physical meaning and would be completely overlooked by the WOM.
However, for the particular case of photoionization from a metallic surface,
in the limit of $a\rightarrow \infty $ Ramsauer-Townsend oscillations must
disappear.

\begin{figure}[!htb]
\centering 
\includegraphics[width=0.7\columnwidth]{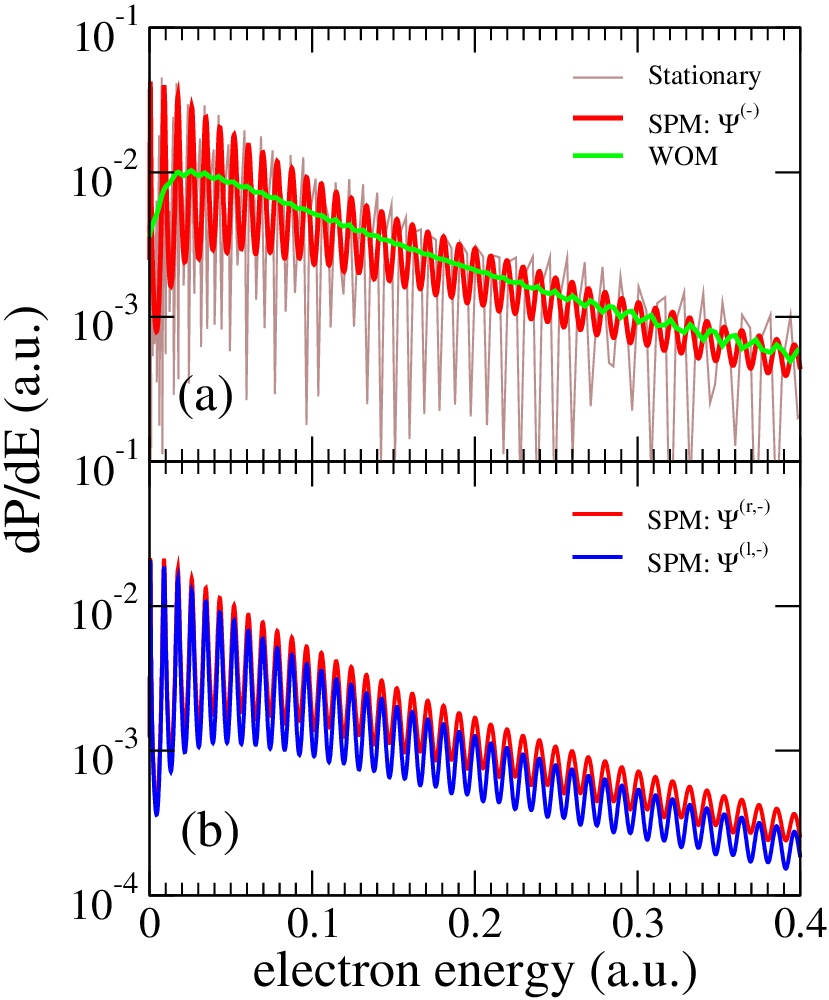} 
\caption{Photoelectron spectrum perpendicular to the surface for
Al (111) within the jellium model: $\Phi =0.156$ a.u., $E_{F}=0.414$ a.u., $%
a=400$ a.u., $V_{0}=0.57$ a.u. subject to a half-cycle pulse of duration $%
\tau =1$ a.u. and momentum transfer $\Delta p=-0.05$ a.u.. (a) Projection
onto eigenfunctions of the TISE with definite parity in brown thin solid
line (highly oscillating), projection onto outgoiing states $\Psi ^{(-)}$ in
red solid line and WOM spectrum in thick green line ($\gamma =0.01$ a.u. and 
$n=3$). (b) Directional spectra calculated by projecting the final wave
function onto $\Psi ^{(r,+)}$ and $\Psi ^{(l,+)}$.}
\label{fig:alhalf}
\end{figure}

When a half-cycle pulse of duration $\tau =1$ and momentum transfer $\Delta
p=-0.05$ a.u. is applied to the Al(111) surface modeled with the jellium
model, we observe in Fig. \ref{fig:alhalf}a that the highly oscillating total spectrum
calculated by projecting onto the stationary states of the continuum is
completely smoothed by the WOM using a window width $\gamma =0.0045$ a.u.
and $n=3.$ However, the SPM accounts for the Ramsauer-Townsend oscillations
with high precision. In Fig. \ref{fig:alhalf}b, we show emission in both directions
calculated within the SPM, also displaying Ramsauer-Towsend resonances.
The emission to the right (left) is the $57\%$ ($43\%$) of a very low ionization
probability $P_{\mathrm{ion}}=2\times 10^{-5}.$

\subsection{Band-Structure-Based surface model}

\begin{figure}[!htb]
\centering 
\includegraphics[width=0.7\columnwidth]{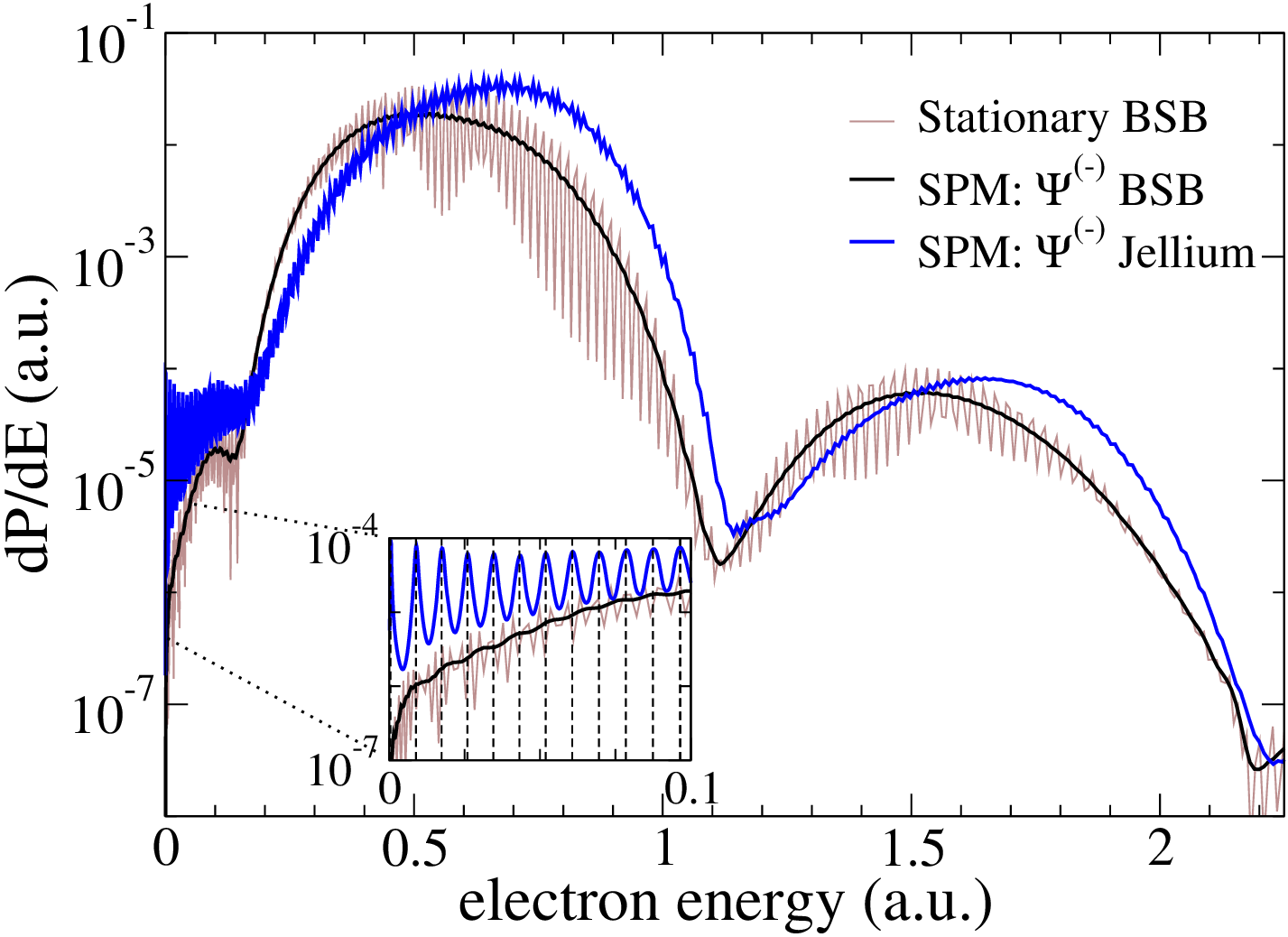} 
\caption{Photoelectron spectrum perpendicular to the surface
Al(111) within the BSB model: $\Phi =0.156$ a.u., $E_{F}=0.414$ a.u., and $%
a=311$ a.u. subject to the same laser pulse of Fig. \ref{fig:aljellium}. Projection onto
stationary states with definite parity in brown line (highly oscillating)
and onto outgoing scattering waves $\Psi ^{(-)}.$ We compare the results with
the projection onto scattering waves of the spectrum corresponding to the
jellium model. The inset shows that the Ramsauer-Townsend resonances present
in the jellium model almost vanish in the BSB model.}
\label{fig:albsb}
\end{figure}

Finally, we solve the problem of photoionization of a metal surface in which
case the potential $V_{S}(z)$ in Eq. (\ref{eq:Hamilt}) is replaced by the
more realistic Band-Structure-Based (BSB) potential where the surface is
represented as a collection of atomic planes that together form a \textit{%
slab }and takes into account effects of the band structure of the metal \cite%
{Chulkov99}. The BSB potential is illustrated in Fig. \ref{fig:pot} for the case of
Al(111). We solve Eq. (\ref{TDSE}) using the BSB potential for an Al(111)
surface and calculate the perpendicular photoemission probability produced
by the same laser pulse of Fig. \ref{fig:pot}. A grid of length $L=789.84$ a.u., spacing $\Delta z=0.05$ a.u. with time step $\Delta t=0.000405$ a.u. is used. After
diagonalization \cite{Lapack10} of the BSB potential, we obtain $120$ bound
eigenstates, from which $91$ are occupied below the Fermi level ($E<E_{F}$),
and $467$ continuum eigenstates. In Fig. \ref{fig:albsb}, we compare the total emission
calculated with the projection onto stationary eigenstates using Eq. (\ref%
{spectrum-box}) and the SPM. Unlike square-well potentials, in the case of
the BSB potential the scattering waves do not have any analytical solution
and are unknown \cite{Pang06}. In consequence, in order to implement the SPM
we must appeal to the procedure described in the appendix \ref{appendixA}
(from step 1 to step 7) to obtain the numerically converged outgoing
scattering basis $\left\{ \Psi _{k}^{(-)}(z)\right\} $ computed using a
minimization tolerance of $10^{-12}\mathrm{.}$ Again, spurious oscillations
are correctly eliminated within the SPM. When comparing the total energy
spectrum calculated within the two models: BSB and jellium, two important things are worth mentioning. First, there is a shift of about $0.15$ a.u. in
the position of the peaks, which is the BSB spectrum displaced towards the
threshold with respect to the jellium spectrum. This shift was adjudicated
to the rugosity of the BSB potential \cite{Rios15,Rios12,Rios17}. Very
importantly, the Ramsauer-Towsend oscillations in the jellium spectrum are
washed out in the BSB model because the surface potential is described with
smooth walls.

\section{Conclusions}

\label{conclusions}

We have introduced the scattering projection method (SPM) to retrieve the energy spectrum in one-dimensional photoionization problems. SPM is based on solving the time-dependent Schrödinger equation, and projected the final time-evolved wavefunction onto scattering waves that have appropriate incoming or outgoing boundary conditions. 
This approach was developed to eliminate spurious oscillations that often appear in the ionization spectrum when using a numerical basis generated on a finite grid. We compared our method to another commonly used technique for reducing spurious oscillations, the well-known window-operator method (WOM), and demonstrated several advantages of SPM. 
On one hand, SPM more effectively smooths high-frequency oscillations compared to WOM. Additionally, SPM enables the computation of directional emission, allowing for the investigation of emission asymmetry. We have also studied this asymmetry using directional half-cycle pulses.

In practical applications, such as the electron emission from metal surfaces excited by the grazing incidence of ultrashort laser pulses, previous calculations faced significant challenges from numerous spurious oscillations. 
Our new results show that SPM can resolve certain physical features in the spectra from metal surfaces while disregarding spurious oscillations. Overall, SPM produces smoother ionization spectra more efficiently than WOM, and, importantly, it does not remove other types of physical oscillations that mainly arise from genuine quantum interference effects.

\ack
This work is supported by PICT Grants No. 2020-01755, No. 2020-01434, and 
No. 2020–01931 of ANPCyT (Argentina), 
PIP Grant No. 2022-2024 No 11220210100468CO and No. 11220200102421CO
of Consejo Nacional de Investigaciones Cient\'{\i}ficas y T\'{e}cnicas (CONICET)
(Argentina).

\appendix
\section{Numerical calculation of the scattering waves}
\label{appendixA}

We assume a localized one-dimensional potential ($V(z)\neq 0$ for $%
z_{0}<z<z_{1}$) and an incident particle from the left. Since in all our
cases the potential is asymptotically zero in the external region, the wave
function $\Psi _{k}^{(l,-)}(z)$ in Eq. (\ref{+l}) can be divided in three
different regions: 
\begin{equation}
\Psi _{k}^{(l,-)}(z)=\left\{ 
\begin{array}{lll}
\phi _{I}(z) & \mathrm{if} & z<z_{0} \\ 
\phi _{II}(z) & \mathrm{if} & z_{0}<z<z_{1} \\ 
\phi _{III}(z) & \mathrm{if} & z>z_{1}.%
\end{array}%
\right.  \label{piecewise}
\end{equation}%
First, the transmission $\left\vert T_{l}^{(-)}\right\vert ^{2}$ and
reflection $\left\vert R_{l}^{(-)}\right\vert ^{2}$ coefficients must
satisfy $\left\vert T_{l}^{(-)}\right\vert ^{2}+\left\vert
R_{l}^{(-)}\right\vert ^{2}=1$. Then, the boundary conditions at the edges $%
z_{0}$ and $z_{1}$ are established, requiring the continuity of the wave
function and its derivative. The proposed procedure to obtain $\Psi
_{k}^{(l,-)}(z)$ is the following \cite{Pang06}:

\begin{enumerate}
\item Given the energy of the incident particle $E$, an initial value of the
complex reflection parameter $R_{l}^{(-)}$ is proposed, which must be
restricted to a unit circle because $\left\vert R_{l}^{(+)}\right\vert
^{2}\leq 1$.

\item Applying the continuity criteria, in Eq. (\ref{piecewise}) we use the
function $\phi _{I}(z)$ to determine the function $\phi _{II}(z)$ and its
derivative at the point $z_{0}$.

\item The time-independent Schr\"{o}dinger equation is solved in the
intermediate region by integrating from $z_{0}$ to $z_{1}$ using the 4th
order Runge-Kutta method \cite{Press07}.

\item Knowing the function $\phi _{II}(z_{1})$, and applying the continuity
conditions to Eq. (\ref{piecewise}), the right-hand side function $\phi
_{III}(z)$ is obtained. With it, we get the transmission coefficient given
by $T_{l}^{(-)}(k)=\phi _{III}(z_{1})\,e^{ikz_{1}}$.

\item Using this value of the transmission coefficient $T_{l}^{(+)}(k)$, the
equation is integrated in the opposite direction (from $z_{1}$ to $z_{0}$),
obtaining, in the central region of Eq. (\ref{piecewise}), a new $\phi
_{II}(z)$.

\item On the left edge $z_{0}$, the continuity conditions are applied,
determining $\phi _{I}(z_{0})$. This allows obtaining a new complex
reflection coefficient $\left. R_{l}^{(-)}\right\vert _{\mathrm{new}}=\left[
\phi _{II}(z_{0})-e^{ikz_{0}}\right] \,e^{ikz_{0}}$ from Eq. (\ref{-l}). We
define a real function that corresponds to the Euclidean distance between
the old and new reflection coefficients $R_{l}^{(-)}$ and $\left.
R_{l}^{(-)}\right\vert _{\mathrm{new}}$.

\item An iteration process is performed, such that it minimizes the distance
between the old and new reflection coefficients $R_{l}^{(-)}$ and $\left.
R_{l}^{(-)}\right\vert _{\mathrm{new}}$. The integration problem is now
transformed into an optimization problem. In the present work, this is done
using the \textit{steepest descent} method \cite{Press07}.
\end{enumerate}

In a completely analogous way, we solved the problem of an incident particle
from the right, in which case the scattering wave function $\Psi
_{k}^{(r,-)}(z)$ is obtained. The combination of both types of functions,
calculated at different energies, constitutes a complete scattering basis.
We proceed analogously with $\Psi _{k}^{(l,+)}$ and $\Psi _{k}^{(r,+)}.$

\bibliography{biblio}

\begin{thebibliography}{10}

\bibitem{Nobel2023}
Peter {Dombi} and Martin {Schultze}.
\newblock {The Nobel Prize in Physics 2023}.
\newblock {\em Europhysics News}, 54(5):8--9, December 2023.

\bibitem{Agostini79}
P.~{Agostini}, F.~{Fabre}, G.~{Mainfray}, G.~{Petite}, and N.~K. {Rahman}.
\newblock {Free-Free Transitions Following Six-Photon Ionization of Xenon Atoms}.
\newblock {\em Physical Review Letters}, 42:1127--1130, April 1979.

\bibitem{DiMauro95}
L.~F. {Dimauro} and P.~{Agostini}.
\newblock {Ionization Dynamics in Strong Laser Fields}.
\newblock {\em Advances in Atomic Molecular and Optical Physics}, 35:79--120, January 1995.

\bibitem{Schultze10}
M.~{Schultze}, M.~{Fie{\ss}}, N.~{Karpowicz}, J.~{Gagnon}, M.~{Korbman}, M.~{Hofstetter}, S.~{Neppl}, A.~L. {Cavalieri}, Y.~{Komninos}, Th. {Mercouris}, C.~A. {Nicolaides}, R.~{Pazourek}, S.~{Nagele}, J.~{Feist}, J.~{Burgd{\"o}rfer}, A.~M. {Azzeer}, R.~{Ernstorfer}, R.~{Kienberger}, U.~{Kleineberg}, E.~{Goulielmakis}, F.~{Krausz}, and V.~S. {Yakovlev}.
\newblock {Delay in Photoemission}.
\newblock {\em Science}, 328(5986):1658, June 2010.

\bibitem{Krausz09}
Ferenc Krausz and Misha Ivanov.
\newblock Attosecond physics.
\newblock {\em Rev. Mod. Phys.}, 81:163--234, Feb 2009.

\bibitem{Dutoi11}
Anthony~D. {Dutoi}, Michael {Wormit}, and Lorenz~S. {Cederbaum}.
\newblock {Ultrafast charge separation driven by differential particle and hole mobilities}.
\newblock {\em \jcp}, 134(2):024303--024303, January 2011.

\bibitem{Huppert16}
Martin {Huppert}, Inga {Jordan}, Denitsa {Baykusheva}, Aaron {von Conta}, and Hans~Jakob {W{\"o}rner}.
\newblock {Attosecond Delays in Molecular Photoionization}.
\newblock {\em \prl}, 117(9):093001, August 2016.

\bibitem{Ning14}
Qi-Cheng {Ning}, Liang-You {Peng}, Shu-Na {Song}, Wei-Chao {Jiang}, Stefan {Nagele}, Renate {Pazourek}, Joachim {Burgd{\"o}rfer}, and Qihuang {Gong}.
\newblock {Attosecond streaking of Cohen-Fano interferences in the photoionization of H2+}.
\newblock {\em \pra}, 90(1):013423, July 2014.

\bibitem{Cavalieri07}
A.~L. {Cavalieri}, N.~{M{\"u}ller}, Th. {Uphues}, V.~S. {Yakovlev}, A.~{Baltu{\v{s}}ka}, B.~{Horvath}, B.~{Schmidt}, L.~{Bl{\"u}mel}, R.~{Holzwarth}, S.~{Hendel}, M.~{Drescher}, U.~{Kleineberg}, P.~M. {Echenique}, R.~{Kienberger}, F.~{Krausz}, and U.~{Heinzmann}.
\newblock {Attosecond spectroscopy in condensed matter}.
\newblock {\em \nat}, 449(7165):1029--1032, October 2007.

\bibitem{Neppl12}
S.~Neppl, R.~Ernstorfer, E.~M. Bothschafter, A.~L. Cavalieri, D.~Menzel, J.~V. Barth, F.~Krausz, R.~Kienberger, and P.~Feulner.
\newblock Attosecond time-resolved photoemission from core and valence states of magnesium.
\newblock {\em Phys. Rev. Lett.}, 109:087401, Aug 2012.

\bibitem{Ossiander18}
M.~{Ossiander}, J.~{Riemensberger}, S.~{Neppl}, M.~{Mittermair}, M.~{Sch{\"a}ffer}, A.~{Duensing}, M.~S. {Wagner}, R.~{Heider}, M.~{Wurzer}, M.~{Gerl}, M.~{Schnitzenbaumer}, J.~V. {Barth}, F.~{Libisch}, C.~{Lemell}, J.~{Burgd{\"o}rfer}, P.~{Feulner}, and R.~{Kienberger}.
\newblock {Absolute timing of the photoelectric effect}.
\newblock {\em \nat}, 561(7723):374--377, September 2018.

\bibitem{Rios12}
C.~A. {Rios Rubiano}, M.~S. {Gravielle}, D.~M. {Mitnik}, and V.~M. {Silkin}.
\newblock {Band-structure effects in photoelectron-emission spectra from metal surfaces}.
\newblock {\em \pra}, 85(4):043422, April 2012.

\bibitem{Rios17}
C.~A. {R{\'\i}os Rubiano}, R.~{Della Picca}, D.~M. {Mitnik}, V.~M. {Silkin}, and M.~S. {Gravielle}.
\newblock {Induced-field enhancement of band-structure effects in photoelectron spectra from Al surfaces by ultrashort laser pulses}.
\newblock {\em \pra}, 95(3):033401, March 2017.

\bibitem{Faraggi07}
M.~N. {Faraggi}, M.~S. {Gravielle}, and D.~M. {Mitnik}.
\newblock {Interaction of ultrashort laser pulses with metal surfaces: Impulsive jellium-Volkov approximation versus the solution of the time-dependent Schr{\"o}dinger equation}.
\newblock {\em \pra}, 76(1):012903, July 2007.

\bibitem{Fetic20}
B.~{Feti{\'c}}, W.~{Becker}, and D.~B. {Milo{\v{s}}evi{\'c}}.
\newblock {Extracting photoelectron spectra from the time-dependent wave function: Comparison of the projection onto continuum states and window-operator methods}.
\newblock {\em \pra}, 102(2):023101, August 2020.

\bibitem{Schafer90}
K.~J. {Schafer} and K.~C. {Kulander}.
\newblock {Energy analysis of time-dependent wave functions: Application to above-threshold ionization}.
\newblock {\em \pra}, 42(9):5794--5797, November 1990.

\bibitem{Schafer91}
K.~J. {Schafer}.
\newblock {The energy analysis of time-dependent numerical wave functions}.
\newblock {\em Computer Physics Communications}, 63(1-3):427--434, February 1991.

\bibitem{Chulkov99}
E.~V. {Chulkov}, V.~M. {Silkin}, and P.~M. {Echenique}.
\newblock {Image potential states on metal surfaces: binding energies and wave functions}.
\newblock {\em Surface Science}, 437(3):330--352, September 1999.

\bibitem{Moler03}
Cleve {Moler} and Charles {van Loan}.
\newblock {Nineteen Dubious Ways to Compute the Exponential of a Matrix, Twenty-Five Years Later}.
\newblock {\em SIAM Review}, 45(1):3--49, January 2003.

\bibitem{Press07}
William~H. Press, Saul~A. Teukolsky, William~T. Vetterling, and Brian~P. Flannery.
\newblock {\em Numerical Recipes 3rd Edition: The Art of Scientific Computing}.
\newblock Cambridge University Press, 3 edition, 2007.

\bibitem{Pindzola07}
M.~S. {Pindzola}, F.~{Robicheaux}, S.~D. {Loch}, J.~C. {Berengut}, T.~{Topcu}, J.~{Colgan}, M.~{Foster}, D.~C. {Griffin}, C.~P. {Ballance}, D.~R. {Schultz}, T.~{Minami}, N.~R. {Badnell}, M.~C. {Witthoeft}, D.~R. {Plante}, D.~M. {Mitnik}, J.~A. {Ludlow}, and U.~{Kleiman}.
\newblock {TOPICAL REVIEW: The time-dependent close-coupling method for atomic and molecular collision processes}.
\newblock {\em Journal of Physics B Atomic Molecular Physics}, 40(7):R39--R60, April 2007.

\bibitem{Griffiths18}
David~J. Griffiths and Darrell~F. Schroeter.
\newblock {\em Introduction to quantum mechanics}.
\newblock Cambridge University Press, Cambridge ; New York, NY, third edition edition, 2018.

\bibitem{Pang06}
Tao Pang.
\newblock {\em An introduction to computational physics}.
\newblock Cambridge University Press,, Cambridge ;, 2nd ed. edition, 2006.

\bibitem{Garriz10}
Abel~E. {Garriz}, Alejandro {Sztrajman}, and Dar{\'\i}o {Mitnik}.
\newblock {Running into trouble with the time-dependent propagation of a wavepacket}.
\newblock {\em European Journal of Physics}, 31(4):785--799, July 2010.

\bibitem{Lapack10}
National~Science Foundation and Department of~Energy.
\newblock {LAPACK} -- linear algebra {PACKage}.
\newblock http://www.netlib.org/lapack/, 2010.

\bibitem{Arbo08a}
D.~G. {Arb{\'o}}, J.~E. {Miraglia}, M.~S. {Gravielle}, K.~{Schiessl}, E.~{Persson}, and J.~{Burgd{\"o}rfer}.
\newblock {Coulomb-Volkov approximation for near-threshold ionization by short laser pulses}.
\newblock {\em Physical Review A}, 77(1):013401, January 2008.

\bibitem{Lancaster03}
C.~L. {Stokely}, J.~C. {Lancaster}, F.~B. {Dunning}, D.~G. {Arb{\'o}}, C.~O. {Reinhold}, and J.~{Burgd{\"o}rfer}.
\newblock {Production of quasi-one-dimensional very-high-n Rydberg atoms}.
\newblock {\em \pra}, 67(1):013403, January 2003.

\bibitem{Arbo03}
D.~G. {Arb{\'o}}, C.~O. {Reinhold}, J.~{Burgd{\"o}rfer}, A.~K. {Pattanayak}, C.~L. {Stokely}, W.~{Zhao}, J.~C. {Lancaster}, and F.~B. {Dunning}.
\newblock {Pulse-induced focusing of Rydberg wave packets}.
\newblock {\em \pra}, 67(6):063401, June 2003.

\bibitem{Rios15}
C.~A. {R{\'\i}os Rubiano}.
\newblock {\em Interacci\'on din\'amica r\'apida de part{\'\i}culas y campos electromagn\'eticos con superficies met\'alicas}.
\newblock PhD thesis, Universidad de Buenos Aires - CONICET. Instituto de Astronom{\'\i}a y F{\'\i}sica del Espacio (IAFE). Grupo de Din\'amica Cu\'antica en la Materia, https://bibliotecadigital.exactas.uba.ar/collection/tesis/document/tesis\_n5685\_RiosRubiano, 2015.

\end{thebibliography}

\end{document}